\documentclass[showpacs,aps,prl,twocolumn,superscriptaddress]{revtex4}
\usepackage{graphicx} 
\usepackage{dcolumn}
\usepackage{bm}

\begin{document}

\title{Chirality-Selective Excitation of Coherent Phonons in Carbon Nanotubes by Femtosecond Optical Pulses}

\author{J.-H.~Kim}
\author{K.-J.~Han}
\author{N.-J.~Kim}
\affiliation{Department of Physics, Chungnam National University, Daejeon, 305-764, Republic of Korea}

\author{K.-J.~Yee}
\email[]{kyee@cnu.ac.kr}
\thanks{corresponding author.}
\affiliation{Department of Physics, Chungnam National University, Daejeon, 305-764, Republic of Korea}

\author{Y.-S.~Lim}
\affiliation{Department of Applied Physics, Konkuk University, Chungju, Chungbuk, 380-701, Republic of Korea}

\author{G.~D.~Sanders}
\author{C.~J.~Stanton}
\affiliation{Department of Physics, University of Florida, Gainesville, Florida 32611, USA}

\author{L.~G.~Booshehri}
\author{E.~H.~H\'{a}roz}
\affiliation{Department of Electrical and Computer Engineering, Rice University, Houston, Texas 77005, USA}

\author{J.~Kono}
\email[]{kono@rice.edu}
\thanks{corresponding author.}
\affiliation{Department of Electrical and Computer Engineering, Rice University, Houston, Texas 77005, USA}

\date{\today}

\begin{abstract}
Using pre-designed trains of femtosecond optical pulses, we have selectively excited coherent phonons of the radial breathing mode of specific-chirality single-walled carbon nanotubes within an ensemble sample.  By analyzing the initial phase of the phonon oscillations, we prove that the tube diameter initially increases in response to ultrafast photoexcitation.  Furthermore, from excitation profiles, we demonstrate that an excitonic absorption peak of carbon nanotubes periodically oscillates as a function of time when the tube diameter undergoes radial breathing mode oscillations.
\end{abstract}

\pacs{78.67.Ch,71.35.Ji,78.55.-m}

\maketitle



Single-walled carbon nanotubes (SWNTs), hollow one-dimensional nanostructures with unique electronic, mechanical, and
optical properties, come in a variety of species, or chiralities.  Some of them are metallic and others semiconducting,
depending on their chiral indices ($n$,$m$)~\cite{Dresselhaus08Book,Oconnell06Book,DresselhausetAl01Book}.  This
diversity, while making them such unusual nanomaterials, often makes it challenging to extract reliable parameters on
chirality-dependent properties from experimental results on ensemble samples.  Currently, there are world-wide efforts
on SWNT purification, separation, and enrichment, producing promising
results~\cite{BachiloetAl03NL,ArnoldetAl05NL,YangetAl05JPCB,MiyataetAl06JPCB,ArnoldetAl06NatureNano}.
However, a standard for fabrication of these samples has yet to be established.

\begin{figure}
\includegraphics [scale=0.41] {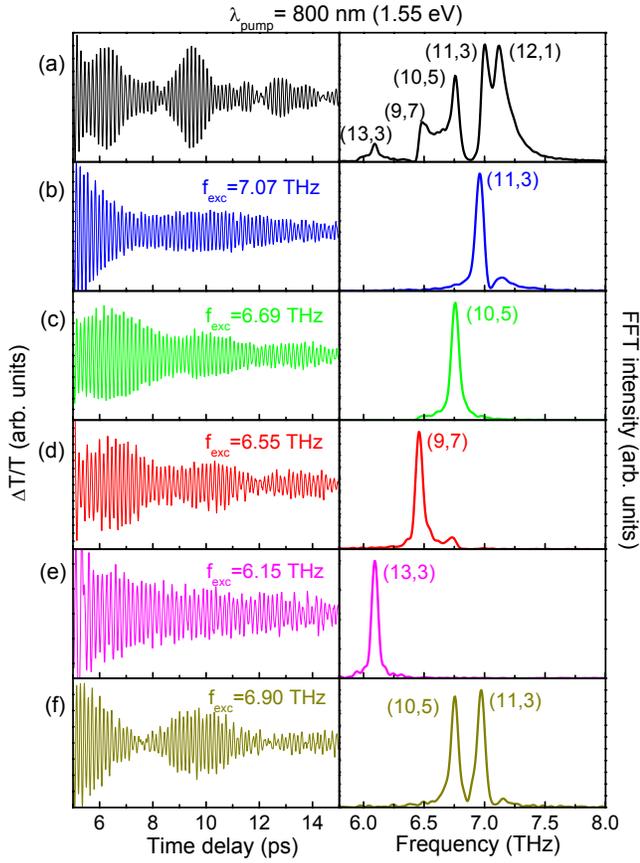}
\caption{(color online).  (a): (Left) Time-domain transmission modulations due to coherent RBM vibrations in
ensemble SWNT solution generated using standard pump-probe techniques {\em without} pulse shaping;
(Right) Fourier transformation of time-domain oscillations with chirality assigned peaks.
(b-f): (Left) Time-domain coherent RBM oscillations selectively excited by multiple pulse trains via pulse
shaping with corresponding repetition rates of 7.07-6.15~THz; (Right) Fourier transformations of corresponding
oscillations, with their dominant nanotube chirality ($n$,$m$) indicated.  In (f), the pump repetition rate was
intentionally tuned to the middle of the RBM frequencies for (11,3) and (10,5) nanotubes.}
\label{cp}
\end{figure}

Here, we present a novel method that allows us to study single-chirality nanotubes even though the sample
contains nanotubes of many different chiralities.  Specifically, we have utilized the techniques of femtosecond
pulse shaping~\cite{WeineretAl88JOSAB,WeinerLearird90OL,LaarmannetAl07PRL} in ultrafast pump-probe spectroscopy
of SWNTs to selectively excite the coherent lattice vibrations~\cite{LimetAl06NL,GambettaetAl06NP} of the radial
breathing mode (RBM) of specific chiralities.  The excitation spectra of such coherent phonons (CPs) provide
chirality-specific information on the processes of light absorption, phonon generation, and phonon-induced band
structure modulations in unprecedented detail.  
In particular, 
the excitation-energy-dependence of the phase of the CP oscillations provides direct, time-domain evidence that band gap oscillations follow the diameter oscillations in the RBM coherent phonon mode.

The sample studied was a micelle-suspended SWNT solution, where the SWNTs (HiPco batch HPR 104) were suspended
as individuals with sodium cholate~\cite{OconnelletAl02Science}. The optical setup was that of standard degenerate
pump-probe spectroscopy, but chirality selectivity was achieved by using multiple pulse trains, with a pulse-to-pulse
interval corresponding to the period of a specific RBM mode. Among different species of nanotubes, those having
RBM frequencies that are matched to the repetition rate of multiple pulse trains will generate large amplitude
coherent oscillations with increasing oscillatory response to each pulse, while others will have diminished
coherent responses~\cite{WeineretAl90Science,LeeetAl06PRB,HaseetAl96APL}.  The tailoring of multiple pulse trains
from femtosecond pulses was achieved using the pulse-shaping technique described elsewhere~\cite{WeinerLearird90OL}.
Pulse trains are incident on an ensemble of nanotubes as a pump beam, and coherent RBM oscillations are monitored
by an unshaped, Gaussian probe beam.  The pump pulse fluence was around 4 $\times$ 10$^{-6}$~J/cm$^2$, where no
noticeable change in the RBM frequency or in the decay rate was observed when the pump fluence was adjusted.

The strength of our pulse-shaping technique is shown in Fig.~1.  Real-time observation of coherent RBM oscillations
is possible without pulse-shaping by employing standard femtosecond pump-probe
spectroscopy~\cite{LimetAl06NL,GambettaetAl06NP}.  Figure 1(a) shows transmission modulations of the probe beam
induced by coherent phonon lattice vibrations, which were generated by pump pulses with a pulse-width of 50~fs
and a central wavelength of 800~nm.  The time-domain beating profiles reflect the simultaneous generation of
several RBM frequencies~\cite{LimetAl06NL}, which are clearly seen in the Fourier-transform of the time-domain data in the right of Fig.~1(a).  Although resonance conditions and RBM frequencies lead to the assignment of peaks to their
corresponding chiralities, obtaining detailed information on dynamical quantities such as the phase information
of phonon oscillations becomes rather challenging.  Additionally, if adjacent phonon modes overlap in the spectral
domain, this can lead to peak distortions.

However, by introducing pulse-shaping, multiple pulses with different repetition rates are used to excite
RBM oscillations, and, as shown in Figs.~1(b)-1(e), chirality selectivity was successfully obtained.  With the
appropriate repetition rate of the pulse trains, a single, specific chirality dominantly contributes to the
signal, while other nanotubes are suppressed.  For example, by choosing a pump repetition rate of 7.07~THz, we
can selectively excite only the (11,3) nanotubes, as seen in Fig.~1(b).  Similarly, with a pump repetition
rate of 6.69~THz, the (10,5) nanotubes are selectively excited, as seen in Fig.~1(c).  Finally, as demonstrated
in Fig.~1(f), when the pump repetition rate was tuned to the middle of the RBM frequencies for (11,3) and (10,5)
nanotubes, both nanotubes contributed.  The accuracy of selectivity depends on the number of pulses in the
tailored pulse train as well as on the distribution of chiralities in the nanotube ensemble.  Furthermore,
selective excitation of a specific chirality also requires the pump energy to be resonant with the corresponding
second interband (or $E_{22}$) transition.

\begin{figure}
\includegraphics [scale=0.73]{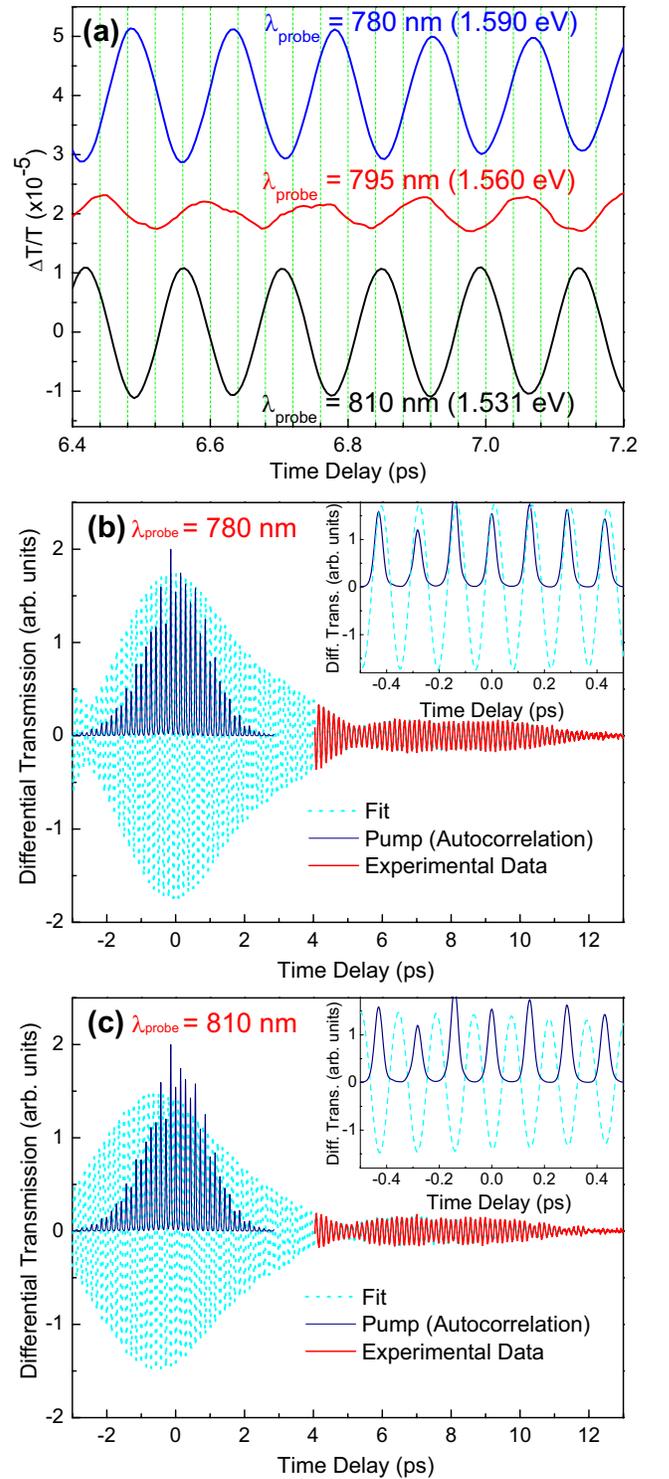}
\caption{(color online).  (a) Differential transmission as a function of time delay at probe wavelengths
of 780~nm, 795~nm, and 810~nm for the selective RBM excitation of the (11,3) nanotubes.  There is a $\pi$ phase
shift between the 780~nm and 810~nm data.  These three wavelengths correspond to below, at, and above the second
exciton resonance, respectively, of (11,3) nanotubes.  (b, c): Differential transmission as a function of time
delay together with the pump pulse train as well as fit to exponentially-decaying sinusoidal oscillations.}
\label{figBdep}
\end{figure}
The ability to excite single-chirality tubes allows us to perform detailed studies of their excited states.  For example, by placing a series of 10-nm band pass filters in the probe path before the detector, we can measure the wavelength-dependence of RBM-induced transmission changes to understand how the diameter change during RBM oscillations modifies the band structure.  Figure 2(a) shows the differential transmission for three cases, corresponding to wavelengths below resonance, at resonance, and above resonance, respectively, for selectively-excited (11,3) nanotubes.
Although the transmission is strongly modulated at the RBM frequency (6.96~THz) for all three cases, the amplitude
and phase of oscillations vary noticeably for varying probe wavelengths.  Specifically, the amplitude of oscillations
becomes minimal at resonance, and, in addition, there is clearly a $\pi$-phase shift between the above- and
below-resonance traces.  Because the band gap energy and diameter are inversely related to each other in
SWNTs~\cite{DresselhausetAl01Book}, and because it is the RBM frequency at which the diameter is oscillating, we
can conclude from this data that the energy of the $E_{22}$ resonance is oscillating at the RBM frequency.
Namely, when the band gap is decreasing, absorption above (below) resonance is decreasing (increasing), resulting
in positive (negative) differential transmission.

We can also look at the short time response to see how the diameter changes in response to ultrafast photoexcitation
of electron-hole pairs.  In Fig.~2(b) and 2(c), we plot the differential transmission data taken at probe wavelengths
above (a) and below (b) resonance together with the pump pulse train, with time zero corresponding to the center of
the pulse train.  The sign of the differential transmission oscillations in the first-quarter period is positive
(negative) for the above (below) resonance probe, indicating that there is an initial decrease (increase) in
absorption for above (below) resonance.  Thus, this observation has an important implication for the nature of
coherent phonon RBM diameter oscillations: the lattice must initially expand.  This is an experimental observation
that until now has not been verified, though theoretically predicted~\cite{DumitricaetAl06PRB}.

\begin{figure}
\includegraphics [scale=0.72] {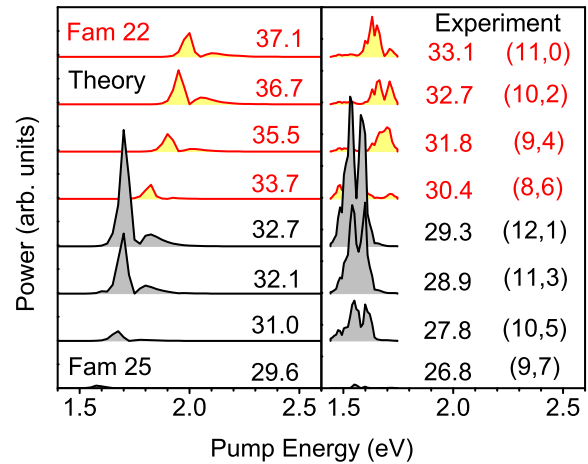}
\caption{(color online) Coherent phonon intensity at the RBM frequency as a function of pump/probe energy for several mod 2 semiconducting nanotubes at the $E_{22}$ transition. Right: experimental spectra; left: simulated spectra. The upper four curves in each panel are for nanotubes in Family 22 and the lower four curves are for tubes in Family 25.
Each curve is labeled with the chirality ($n$,$m$) and the RBM phonon energy in meV and offset for clarity.}
\label{figdeltax}
\end{figure}

Our observations are supported by a microscopic theory we developed for the generation and detection of CP in SWNTs
using an extended tight-binding model~\cite{Porezag95.12947} to describe the nanotube electronic states and a valence
force field model~\cite{Jiang06.235434} to describe the phonons.
We have examined trends in the CP spectra within and between mod~2 semiconducting nanotube families by plotting
the theoretical CP intensity at the RBM phonon frequency as a function of pump/probe energy. This is done in the
left panel of Fig.~3 where we plot our theoretical CP intensity at the RBM frequency as a function of
pump/probe energy for all nanotubes in $2n+m$ Family 22 and 25. The curves for each nanotube are labeled with the
nanotube chirality $(n,m)$ and the RBM phonon energy in meV. In each nanotube, we see peaks in the CP spectra
corresponding to $E_{22}$ transitions. Within a given $2n+m$ family, the CP intensity tends to decrease as the chiral
angle increases, i.e., as the chirality goes from ($n$,0) zigzag tubes to ($n$,$n$) armchair tubes.
From Fig.~3 we can also see that the theoretical CP intensity increases as we go from Family 22 to Family 25.
The right panel of Fig.~3 shows the corresponding experimental CP spectra for the nanotubes in Families 22 and 25.
Comparing experimental and theoretical curves in Fig.~3, we see that our theory correctly predicts the overall
trends in the CP intensities both within and between families~\cite{SandersetAl08cond-mat}.

\begin{figure}
\includegraphics [scale=0.47] {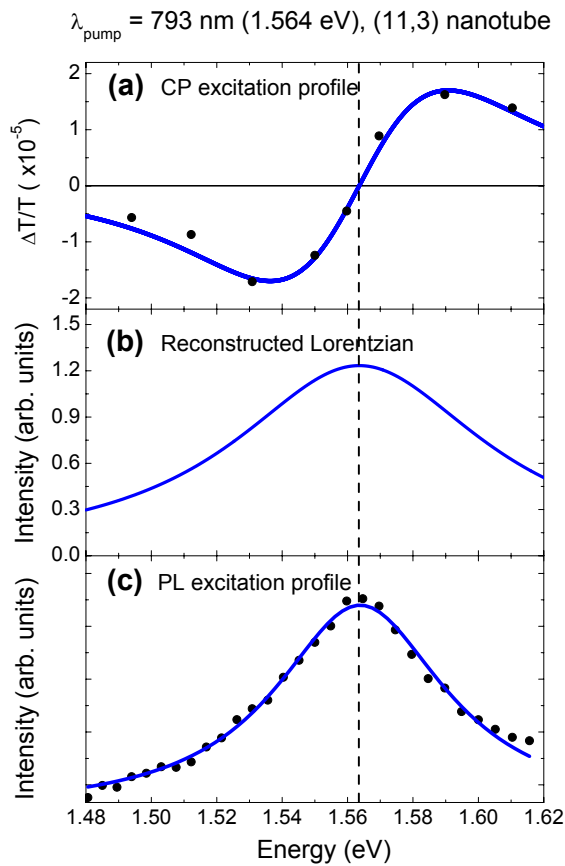}
\caption{(color online).  Excitation profile of coherent phonons as evidence of band gap modulations.
(a) Phonon amplitude as a function of probe energy fit to the derivative of a Lorentzian peak.
(b) Lorentzian reconstructed from (a).  (c) Photoluminescence (PL) excitation profile for (11,3) tubes.}
\label{fig4}
\end{figure}

Considering the phase shift observed in Fig.~2, the amplitude of the differential transmission can be plotted
as a function of probe energy, with the curve of the amplitude fit to the derivative of a Lorentzian function
centered at 1.564~eV as seen in Fig.~4(a).  The fitting to the derivative of a Lorentzian further supports the
idea that modulating the absorption coefficient in time through CP and probing the change in transmission/reflection
at various probe energies is essentially a type of modulation spectroscopy~\cite{Cardona69Book}.  Similar, vibration-induced changes in absorption spectra have been observed in molecular systems~\cite{FragnitoetAl89CPL}.  As compared to previous CP experiments on SWNTs~\cite{LimetAl06NL,GambettaetAl06NP}, pulse-shaping offers phase information that causes the real time-domain data to be sensitive to the sign of the differential transmission, as the amplitude
of the Lorentzian is largest at its inflection points~\cite{LimetAl06NL}.  In Fig.~4(b), the reconstructed
Lorentzian corresponds to the excitation profile for coherent phonons, with a maximum peak comparable to the
maximum position of the photoluminescence excitation profile for (11,3) nanotubes [Fig.~4(c)].

In summary, we produced chirality-selective excitations of coherent RBM oscillations in SWNTs by implementing
multiple pulse trains with repetition rates matched to specific RBM frequencies.  We obtained single-chirality
information for many tube chiralities and extracted detailed information about the phase and modulation of
the absorption, leading to an experimental confirmation that the lattice initially expands in response to
the pump pulse.  Also, the probe energy dependence of the differential transmission is satisfactorily
described by RBM-induced band gap modulations, and analysis of the amplitude further proves coherent phonon
spectroscopy to be a type of modulation spectroscopy.  Beyond providing a powerful tool in accessing
fundamental properties of the material, the technique presented here may lead to chirality-selective end
functionalization and tube filling via CP-driven end cap opening~\cite{DumitricaetAl04PRL,DumitricaetAl06PRB}.

\begin{acknowledgments}
We thank the Korean Research Foundation (MOEHRD, Basic Research Promotion Fund, KRF-2006-311-C00056), the Korea Science and Engineering Foundation (R11-2008-095-01000-0), the Robert A.~Welch Foundation (C-1509), and NSF (DMR-0134058 and DMR-0325474) for support.  We thank J.~Shaver for technical assistance.
\end{acknowledgments}


\begin{thebibliography}{21}
\expandafter\ifx\csname natexlab\endcsname\relax\def\natexlab#1{#1}\fi
\expandafter\ifx\csname bibnamefont\endcsname\relax
  \def\bibnamefont#1{#1}\fi
\expandafter\ifx\csname bibfnamefont\endcsname\relax
  \def\bibfnamefont#1{#1}\fi
\expandafter\ifx\csname citenamefont\endcsname\relax
  \def\citenamefont#1{#1}\fi
\expandafter\ifx\csname url\endcsname\relax
  \def\url#1{\texttt{#1}}\fi
\expandafter\ifx\csname urlprefix\endcsname\relax\def\urlprefix{URL }\fi
\providecommand{\bibinfo}[2]{#2}
\providecommand{\eprint}[2][]{\url{#2}}

\bibitem[{\citenamefont{Jorio et~al.}(2008)\citenamefont{Jorio, Dresselhaus,
  and Dresselhaus}}]{Dresselhaus08Book}
\bibinfo{editor}{\bibfnamefont{A.}~\bibnamefont{Jorio}},
  \bibinfo{editor}{\bibfnamefont{G.}~\bibnamefont{Dresselhaus}},
  \bibnamefont{and} \bibinfo{editor}{\bibfnamefont{M.~S.}
  \bibnamefont{Dresselhaus}}, eds., \emph{\bibinfo{title}{Carbon Nanotubes:
  Advanced Topics in the Synthesis, Structure, Properties and Applications}}
  (\bibinfo{publisher}{Springer}, \bibinfo{address}{Berlin},
  \bibinfo{year}{2008}).

\bibitem[{\citenamefont{O'Connell}(2006)}]{Oconnell06Book}
\bibinfo{editor}{\bibfnamefont{M.~J.} \bibnamefont{O'Connell}}, ed.,
  \emph{\bibinfo{title}{Carbon Nanotubes: Properties and Applications}}
  (\bibinfo{publisher}{CRC Press},
  \bibinfo{address}{Boca Raton}, \bibinfo{year}{2006}).

\bibitem[{\citenamefont{Dresselhaus et~al.}(2001)\citenamefont{Dresselhaus,
  Dresselhaus, and Avouris}}]{DresselhausetAl01Book}
\bibinfo{editor}{\bibfnamefont{M.~S.} \bibnamefont{Dresselhaus}},
  \bibinfo{editor}{\bibfnamefont{G.}~\bibnamefont{Dresselhaus}},
  \bibnamefont{and} \bibinfo{editor}{\bibfnamefont{P.}~\bibnamefont{Avouris}},
  eds., \emph{\bibinfo{title}{Carbon Nanotubes: Synthesis, Structure,
  Properties, and Applications}} (\bibinfo{publisher}{Springer},
  \bibinfo{address}{Berlin}, \bibinfo{year}{2001}).

\bibitem[{\citenamefont{Bachilo et~al.}(2003)\citenamefont{Bachilo, Balzano,
  Herrera, Pompeo, Resasco, and Weisman}}]{BachiloetAl03NL}
\bibinfo{author}{\bibfnamefont{S.~M.} \bibnamefont{Bachilo}}
{\it et al}.,
  \bibinfo{journal}{J. Am. Chem. Soc.}
  \textbf{\bibinfo{volume}{125}}, \bibinfo{pages}{11186}
  (\bibinfo{year}{2003}).

\bibitem[{\citenamefont{Arnold et~al.}(2005)\citenamefont{Arnold, Stupp, and
  Hersam}}]{ArnoldetAl05NL}
\bibinfo{author}{\bibfnamefont{M.~S.} \bibnamefont{Arnold}},
  \bibinfo{author}{\bibfnamefont{S.~I.} \bibnamefont{Stupp}}, \bibnamefont{and}
  \bibinfo{author}{\bibfnamefont{M.~C.} \bibnamefont{Hersam}},
  \bibinfo{journal}{Nano Lett.} \textbf{\bibinfo{volume}{5}},
  \bibinfo{pages}{713} (\bibinfo{year}{2005}).

\bibitem[{\citenamefont{Yang et~al.}(2005)\citenamefont{Yang, Park, An, Lim,
  Seo, Kim, Park, Han, Park, and Lee}}]{YangetAl05JPCB}
\bibinfo{author}{\bibfnamefont{C.-M.} \bibnamefont{Yang}}
{\it et al}.,
  \bibinfo{journal}{J. Phys. Chem. B} \textbf{\bibinfo{volume}{109}},
  \bibinfo{pages}{19242} (\bibinfo{year}{2005}).

\bibitem[{\citenamefont{Miyata et~al.}(2006)\citenamefont{Miyata, Maniwa, and
  Kataura}}]{MiyataetAl06JPCB}
\bibinfo{author}{\bibfnamefont{Y.}~\bibnamefont{Miyata}},
  \bibinfo{author}{\bibfnamefont{Y.}~\bibnamefont{Maniwa}}, \bibnamefont{and}
  \bibinfo{author}{\bibfnamefont{H.}~\bibnamefont{Kataura}},
  \bibinfo{journal}{J. Phys. Chem. B} \textbf{\bibinfo{volume}{110}},
  \bibinfo{pages}{25} (\bibinfo{year}{2006}).

\bibitem[{\citenamefont{Arnold et~al.}(2006)\citenamefont{Arnold, Green,
  Hulvat, Stupp, and Hersam}}]{ArnoldetAl06NatureNano}
\bibinfo{author}{\bibfnamefont{M.~S.} \bibnamefont{Arnold}}
{\it et al}.,
  \bibinfo{journal}{Nature Nanotechnol.} \textbf{\bibinfo{volume}{1}},
  \bibinfo{pages}{60} (\bibinfo{year}{2006}).

\bibitem[{\citenamefont{Weiner et~al.}(1988)\citenamefont{Weiner, Heritage, and
  Kirschner}}]{WeineretAl88JOSAB}
\bibinfo{author}{\bibfnamefont{A.~M.} \bibnamefont{Weiner}},
  \bibinfo{author}{\bibfnamefont{J.~P.} \bibnamefont{Heritage}},
  \bibnamefont{and} \bibinfo{author}{\bibfnamefont{E.~M.}
  \bibnamefont{Kirschner}}, \bibinfo{journal}{J. Opt. Soc. Am. B}
  \textbf{\bibinfo{volume}{5}}, \bibinfo{pages}{1563} (\bibinfo{year}{1988}).

\bibitem[{\citenamefont{Weiner and Leaird}(1990)}]{WeinerLearird90OL}
\bibinfo{author}{\bibfnamefont{A.~M.} \bibnamefont{Weiner}} \bibnamefont{and}
  \bibinfo{author}{\bibfnamefont{D.~E.} \bibnamefont{Leaird}},
  \bibinfo{journal}{Opt. Lett.} \textbf{\bibinfo{volume}{15}},
  \bibinfo{pages}{51} (\bibinfo{year}{1990}).

\bibitem[{\citenamefont{Laarmann et~al.}(2007)\citenamefont{Laarmann,
  Shchatsinin, Stalmashonak, Boyle, Zhavoronkov, Handt, Schmidt, Schulz, and
  Hertel}}]{LaarmannetAl07PRL}
\bibinfo{author}{\bibfnamefont{T.}~\bibnamefont{Laarmann}}
{\it et al}.,
  \bibinfo{journal}{Phys. Rev. Lett.}
  \textbf{\bibinfo{volume}{98}}, \bibinfo{pages}{058302}
  (\bibinfo{year}{2007}).

\bibitem[{\citenamefont{Lim et~al.}(2006)\citenamefont{Lim, Yee, Kim, Haroz,
  Shaver, Kono, Doorn, Hauge, and Smalley}}]{LimetAl06NL}
\bibinfo{author}{\bibfnamefont{Y.~S.} \bibnamefont{Lim}}
{\it et al}.,
  \bibinfo{journal}{Nano Lett.} \textbf{\bibinfo{volume}{6}},
  \bibinfo{pages}{2696} (\bibinfo{year}{2006}).

\bibitem[{\citenamefont{Gambetta et~al.}(2006)\citenamefont{Gambetta, Manzoni,
  Menna, Meneghetti, Cerullo, Lanzani, Tretiak, Piryatinski, Saxena, Martin
  et~al.}}]{GambettaetAl06NP}
\bibinfo{author}{\bibfnamefont{A.}~\bibnamefont{Gambetta}}
{\it et al}.,
  \bibinfo{journal}{Nature Physics}
  \textbf{\bibinfo{volume}{2}}, \bibinfo{pages}{515} (\bibinfo{year}{2006}).

\bibitem[{\citenamefont{O'Connell et~al.}(2002)\citenamefont{O'Connell,
  Bachilo, Huffman, Moore, Strano, Haroz, Rialon, Boul, Noon, Kittrell
  et~al.}}]{OconnelletAl02Science}
\bibinfo{author}{\bibfnamefont{M.~J.} \bibnamefont{O'Connell}}
{\it et al}.,
  \bibinfo{journal}{Science}
  \textbf{\bibinfo{volume}{297}}, \bibinfo{pages}{593} (\bibinfo{year}{2002}).

\bibitem[{\citenamefont{Weiner et~al.}(1990)\citenamefont{Weiner, Leaird,
  Wiederrecht, and Nelson}}]{WeineretAl90Science}
\bibinfo{author}{\bibfnamefont{A.~M.} \bibnamefont{Weiner}}
{\it et al}.,
  \bibinfo{journal}{Science}
  \textbf{\bibinfo{volume}{247}}, \bibinfo{pages}{1317} (\bibinfo{year}{1990}).

\bibitem[{\citenamefont{Lee et~al.}(2006)\citenamefont{Lee, Kim, Yee, and
  Lee}}]{LeeetAl06PRB}
\bibinfo{author}{\bibfnamefont{K.~G.} \bibnamefont{Lee}}
{\it et al}.,
  \bibinfo{journal}{Phys. Rev. B} \textbf{\bibinfo{volume}{74}},
  \bibinfo{pages}{113201} (\bibinfo{year}{2006}).

\bibitem{HaseetAl96APL}
  M.~Hase {\it et al}.,
  Appl.~Phys.~Lett.~{\bf 69}, 2474 (1996).

\bibitem[{\citenamefont{Dumitric\u{a} et~al.}(2006)\citenamefont{Dumitric\u{a},
  Garcia, Jeschke, and Yakobson}}]{DumitricaetAl06PRB}
\bibinfo{author}{\bibfnamefont{T.}~\bibnamefont{Dumitric\u{a}}}
{\it et al}.,
  \bibinfo{journal}{Phys. Rev. B}
  \textbf{\bibinfo{volume}{74}}, \bibinfo{pages}{193406}
  (\bibinfo{year}{2006}).

\bibitem{Porezag95.12947}
  D.~Porezag {\it et al}.,
  Phys.~Rev.~B~{\bf 51}, 12947 (1995).

\bibitem[{\citenamefont{Jiang et~al.}(2006)\citenamefont{Jiang, Tang, Wang, and
  Su}}]{Jiang06.235434}
\bibinfo{author}{\bibfnamefont{J.-W.} \bibnamefont{Jiang}}
{\it et al}.,
  \bibinfo{journal}{Phys. Rev. B} \textbf{\bibinfo{volume}{73}},
  \bibinfo{pages}{235434} (\bibinfo{year}{2006}).

\bibitem{SandersetAl08cond-mat}
For more details, see G.~D.~Sanders {\it et al}., arXiv:0812.1953v1 [cond-mat.mes-hall].

\bibitem[{\citenamefont{Cardona}(1969)}]{Cardona69Book}
\bibinfo{author}{\bibfnamefont{M.}~\bibnamefont{Cardona}},
  \emph{\bibinfo{title}{Modulation Spectroscopy}} (\bibinfo{publisher}{Academic
  Press}, \bibinfo{address}{New York}, \bibinfo{year}{1969}).

\bibitem{FragnitoetAl89CPL}
H.~L. Fragnito, J.-Y. Bigot, P.~C. Becker, and C.~V. Shank, Chem. Phys. Lett. {\bf 160}, 101 (1989).

\bibitem[{\citenamefont{Dumitric\u{a} et~al.}(2004)\citenamefont{Dumitric\u{a},
  Garcia, Jeschke, and Yakobson}}]{DumitricaetAl04PRL}
\bibinfo{author}{\bibfnamefont{T.}~\bibnamefont{Dumitric\u{a}}}
{\it et al}.,
  \bibinfo{journal}{Phys. Rev. Lett.}
  \textbf{\bibinfo{volume}{92}}, \bibinfo{pages}{117401}
  (\bibinfo{year}{2004}).

\end{thebibliography}

\end{document}